# Enhanced thermoelectric properties in hybrid graphene-boron nitride nanoribbons

Kaike Yang,[1,2] Yuanping Chen,[1] Roberto D'Agosta,[2,3] Yuee Xie,[1] Jianxin Zhong,[1] and Angel Rubio[2]

[1] *Laboratory for Quantum Engineering and Micro-Nano Energy Technology and Department of Physics, Xiangtan University, Xiangtan 411105, Hunan, China*

[2] *Nano-Bio Spectroscopy Group and ETSF Scientific Development Centre, Departamento Fisica de Materiales, Universidad del Pais Vasco UPV/EHU, Centro de Fisica de Materiales CSIC-UPV/EHU and DIPC, Avenida Tolosa 72, E-20018 San Sebastian, Spain*

[3] *IKERBASQUE, Basque Foundation for Science, E-48011, Bilbao, Spain*

## ABSTRACT

The thermoelectric properties of hybrid graphene-boron nitride nanoribbons (BCNNRs) are investigated using the non-equilibrium Green's function (NEGF) approach. We find that the thermoelectric figure of merit ($ZT$) can be remarkably enhanced by periodically embedding hexagonal BN (*h*-BN) into graphene nanoribbons (GNRs). Compared to pristine GNRs, the $ZT$ for armchair-edged BCNNRs with width index $3p+2$ is enhanced up to 10~20 times while the $ZT$ of nanoribbons with other widths is enhanced just by 1.5~3 times. As for zigzag-edge nanoribbons, the $ZT$ is enhanced up to 2~3 times. This improvement comes from the combined increase in the Seebeck coefficient and the reduction in the thermal conductivity outweighing the decrease in the electrical conductance. In addition, the effect of component ratio of *h*-BN on the thermoelectric transport properties is discussed. These results qualify BCNNRs as a promising candidate for building outstanding thermoelectric devices.

## I. INTRODUCTION

Thermoelectric materials that can convert heat into electric energy have attracted increasing attention from both theoretical and experimental researches due to potential technological applications.[1-5] The thermoelectric conversion efficiency is measured by the figure of merit:[6-9]

$$ZT = \frac{\sigma S^2 T}{(\kappa_e + \kappa_p)}, \tag{1}$$

where $\sigma$, $S$, $T$, $\kappa_e$ and $\kappa_p$ represent the electrical conductance, Seebeck coefficient, absolute temperature, and the electron and phonon contribution to thermal conductivity, respectively. Clearly, an optimal thermoelectric material should bear high electrical conductance, high Seebeck coefficient and, correspondingly, low thermal conductivity.

Graphene and hexagonal boron nitride ($h$-BN) are single-atomic layer honeycomb materials.[10-19] Because of the quantum confinement effect in dimension, graphene and $h$-BN exhibit very high electrical conductance and Seebeck coefficient.[20-22] However both pristine graphene and pristine $h$-BN are poor thermoelectric materials because their thermal conductivity is also very high.[23-26] Recently, a novel nanomaterial compounded of graphene and $h$-BN has been synthesized by thermal catalytic chemical vapour deposition method.[27,28] This novel hybrid nanostructure possesses physical properties distinct from both pristine graphene and pristine $h$-BN. For example, a tunable electronic energy gap dependent on the concentration of $h$-BN emerges in the two-dimensional hybrid sheet;[29,30] while in the one-dimensional hybrid nanoribbons, it is found that the armchair-edged graphene nanoribbons ($A$-GNRs) embedded in $h$-BN are always semiconducting while zigzag-edged structures become half-metallic;[31,32] the thermal conductivity of hybrid nanoribbon is obviously reduced from the interface scattering.[33] Previous studies did only focus on the electronic and thermal transport, respectively, while the thermoelectric properties of hybrid graphene and $h$-BN (BCN) are left unexplored thus far. Because ordered superlattices can form promising structures for improving thermoelectric performance,[34] it is natural to look into what happens if we construct a periodic structure composed of alternating *h*-BN and graphene.

In this paper, we investigate the thermoelectric properties of BCN nanoribbons (BCNNRs) by using the non-equilibrium Green's function (NEGF) approach.[35-39] It is found that the figure of merit can be remarkably enhanced by periodically embedding *h*-BN into the GNRs. Particularly the $ZT$ is enhanced up to 10~20 times for metallic *A*-GNRs with width index $3p+2$ (where $p$ is a positive integer), while the $ZT$ of nanoribbons with other widths is enhanced about 1.5~3 times. Regarding the zigzag-edge structure, the figure of merit for BCNNRs is improved 2~3 times compared to that of pristine zigzag-edged GNRs (*Z*-GNRs). The origin of this enhancement effect is analyzed: it is found that both the electrical conductance and thermal conductivity decrease, whereas the Seebeck coefficient increases with the periodic number. In addition, the influence of the component ratio of *h*-BN on the thermoelectric transport properties is also discussed. Here, it is shown that the figure of merit of BCNNRs at component ratio near 0.5 has larger value. These results indicate that the BCNNRs could be a very promising candidate for the design of excellent thermoelectric devices.

## II. MODEL AND METHOD

Let us consider the armchair-edged BCNNRs (*A*-BCNNRs) shown in Figs. 1(a) and (b) and zigzag-edge BCNNRs (*Z*-BCNNRs) shown in Figs. 1(c) and (d), respectively. The systems are divided into three parts. The left and right leads composed of GNRs have perfect periodicity along the ribbon axis and the ribbon widths are labeled by $W_A$ and $W_Z$ for armchair- and zigzag-edge, respectively. The central scattering region is the hybridized structure consisting of alternating *h*-BN and graphene supercells. $L_S = (L_{BN} + L_C)$ and $L_S^{'} = (L_{BN}^{'} + L_C^{'})$ are introduced to denote the lengths of the supercell in armchair- and zigzag-edged nanoribbons, respectively, where $L_{BN}$ ($L_{BN}^{'}$) and $L_C$ ($L_C^{'}$) separately represent the lengths of *h*-BN and graphene in the supercell in the longitudinal direction. We want to point out that in Figs. 1(a) and (c) the length of *h*-BN equals to the length of graphene, while their lengths in Figs. 1(b) and (d) are different in the supercell. We use $N$ for describing the number of supercell and the component ratio is defined by $\gamma = L_{BN}/L_S$ in *A*-BCNNRs and $\gamma' = L'_{BN}/L'_S$ in Z-BCNNRs. To investigate the thermoelectric properties of these structures, the NEGF method is employed. For the BCNNRs, the tight-binding Hamiltonian for electrons is written as[40]

$$H = \sum_{\alpha=L,C,R} H_\alpha + \sum_{\beta=L,R} (H^{\beta C} + H^{C\beta})$$
$$= \sum_{\alpha=L,C,R} H_\alpha + \sum_{i,j} t^{LC}_{i,j} (a^{\dagger L}_i a^C_j + h.c.) + \sum_{i,j} t^{RC}_{i,j} (a^{\dagger R}_i a^C_j + h.c.), \quad (2)$$

where $H_\alpha = \sum_i \varepsilon^\alpha_i a^{\dagger \alpha}_i a^\alpha_i + \sum_{i,j} t^\alpha_{i,j} a^{\dagger \alpha}_i a^\alpha_j$, $a^{\dagger \alpha}_i$, $a^\alpha_i$ are the creation and annihilation operators, $\varepsilon^\alpha_i$ is the site energy at the $i-th$ atom, and $t^\alpha_{i,j}$ is the hopping integral for nearest neighbors.[41-43] Based on the Hamiltonian, the retarded Green's function is expressed as

$$G^r(E) = [(E+i0^+)I - H_C - \sum\nolimits_L^r - \sum\nolimits_R^r]^{-1}, \quad (3)$$

where $E$ is the electron energy, $I$ is the identity matrix, $H_C$ is the Hamiltonian of the central scattering region, $\sum\nolimits_\beta^r = H^{C\beta} g^r_\beta H^{\beta C}$ denotes the self-energy of the left (*L*) and right (*R*) regions, and $g^r_\beta$ is the surface Green's function of the leads which can be computed by a recursive iteration technique.[44] Once the retarded Green's function is obtained, we can calculate the electronic transmission coefficient

$$T_e(E) = Tr[G^r(E)\Gamma_L G^a(E)\Gamma_R], \quad (4)$$

where $G^a(\omega) = [G^r(\omega)]^\dagger$ is the advanced Green's function, and $\Gamma_\beta = i(\sum\nolimits_\beta^r - \sum\nolimits_\beta^a)$ describes the interaction between the left or right leads and the central region. By using the transmission coefficient, the electrical conductance, the Seebeck coefficient and the electron contributed thermal conductivity can be calculated separately by

$$\sigma(\mu) = e^2 L_0(\mu,T), \quad (5a)$$

$$S(\mu) = \frac{1}{eT} \frac{L_1(\mu,T)}{L_0(\mu,T)}, \quad (5b)$$

$$\kappa_e(\mu) = \frac{1}{T}[L_2(\mu,T) - \frac{L_1(\mu,T)^2}{L_0(\mu,T)}], \tag{5c}$$

where $L_n(\mu,T) = \frac{2}{h}\int_{-\infty}^{\infty} dE\, T_e(E)\,(E-\mu)^n\,[-\frac{\partial f_e(E,\mu,T)}{\partial E}]$ is the Lorenz function, $\mu$ is chemical potential, $e$ is electric charge of carriers, $f_e$ is the Fermi-Dirac distribution function at temperature $T$, and $h$ is the Planck constant.

On the other hand for thermal transport, according to Refs. [36-38], the phonon-transmission coefficient, $T_p(\omega)$, can be calculated with the same steps as for the electrons. Indeed, one only need to change $E$ into $\omega^2$, $H_C$ into $K_C$, and correspondingly compute the self-energy, where $\omega$ is the vibrational frequency of phonons, and $K_C$ is the spring constant matrix of the central region in Eq. (3).[25,26,38,45,46] Then the phonons contributed thermal conductivity is given by

$$\kappa_p(T) = \frac{\hbar}{2\pi}\int_0^{\infty} \frac{\partial f_p(\omega)}{\partial T} T_p(\omega)\, \omega\, d\omega, \tag{6}$$

where $f_p(\omega)$ is the Bose-Einstein distribution function for heat carriers. We have therefore all the ingredients to calculate the figure of merit.

## III. RESULTS AND DISCUSSION

In Fig. 2, we investigate the thermoelectric transport properties of *A*-BCNNRs with different widths as a function of the number of supercell $N$ for $\gamma = 0.5$ ($L_{BN} = L_C$). Figure 2(a) shows $ZT$ for *A*-BCNNRs with width $W_A = 5$, and the inset plots the ratio $ZT/ZT_0$, where $ZT_0$ is the figure of merit of pristine *A*-GNRs. Each $ZT$ in calculation is taken at the highest figure of merit peak value of hybridized nanoribbons. One can find that when the periodic number $N = 1$, $ZT$ is relatively small. Increasing $N$, $ZT$ gradually increases for various $L_{BN}$ values. However, the enhanced magnitude is different, which can be seen from the inset of Fig. 2(a). At $L_{BN} = 4, 7$ and 8, it is found that the figure of merit for *A*-BCNNRs is enhanced about 10 times compared to that of pristine metallic *A*-GNRs. Interestingly, in the case of $L_{BN} = 5$ and 6, quite higher value $ZT/ZT_0 \approx 15 \sim 20$ is attained. When the periodic number is greater than 6, we find that $ZT/ZT_0$ approaches almost a constant value. The reason for this behavior is easily explained: when the periodic number is sufficiently large, the whole nanoribbons can be approximately regarded as a superlattice structure composed of alternating *h*-BN and graphene; then the transmission of electron and phonon in the *A*-BCNNRs is equivalent to that of the superlattice.

In Figs. 2(b) and (c), the thermoelectric properties of semiconducting *A*-GNRs embedded with *h*-BN are investigated. As a comparison, we plot the figure of merit for *A*-BCNNRs with width $W_A = 6$ and 7. It can be seen that similarly to the case of metallic nanoribbons with width $W_A = 5$, $ZT$ increases with the periodic number $N$ and finally approaches a constant value for different $L_{BN}$. Nevertheless, compared to the perfect semiconducting *A*-GNRs, it is found that the enhanced magnitude for *A*-BCNNRs is just about 1.5~3.0 times larger than $ZT$ for pristine graphene nanoribbons (see the insets). When further increasing the ribbon width, a similar phenomenon is also found in the three categories of nanoribbons (not shown).

To elucidate the origin of the thermoelectric figure of merit enhancement in *A*-BCNNRs, figures 3(a)-(c) respectively depict the electrical conductance ratio ($\sigma/\sigma_0$), thermal conductivity ratio ($\kappa/\kappa_0$) and Seebeck coefficient ratio ($S/S_0$) as a function of periodic number $N$ for the ribbon width $W_A = 5$, and, accordingly, in the insets we plot the same quantities for nanoribbons with $W_A = 6$. The quantities with subscript 0 correspond to the parameters of pristine *A*-GNRs. It can be seen that both the electrical conductance and thermal conductivity ratios decrease with increasing the periodic number, although the attenuating magnitude of them is different. In contrast to the perfect *A*-GNRs, the electrical conductance of *A*-BCNNRs is decreased nearly 2~3 times, while the thermal conductivity is decreased 1~2 times, which suggests that the hybrid structure has strong effects on the electronic conduction. The decreased electrical conductance and thermal conductivity ratios can be attributed to interface scattering which blocks the carriers' transmission. From Fig. 3(c), it is seen that the ratio of Seebeck coefficient increases with the periodic number $N$. The enhanced magnitude of Seebeck coefficient for *A*-BCNNRs with width $W_A = 5$ is about 5 times, which is larger than that of the nanoribbons with width $W_A = 6$ [see the inset of Fig. 3(c)]. Therefore, the enhancement of the figure of merit is mainly due to the increase of the Seebeck coefficient and the corresponding decrease of the thermal conductivity. Together, they outweigh the decrease in the electrical conductance. From Eq. (5b) one can find that the Seebeck coefficient is dependent on the ratio of Lorenz function $L_1/L_0$:

$$L_0(\mu,T) = \frac{2}{h}\int_{-\infty}^{\infty} [-\frac{\partial f_e(E,\mu,T)}{\partial E}] T_e(E)\, dE, \qquad (7a)$$

$$L_1(\mu,T) = \frac{2}{h}\{\int_{-\infty}^{E<\mu} [-\frac{\partial f_e(E,\mu,T)}{\partial E}](E-\mu)T_e(E)\,dE + \int_{E>\mu}^{\infty} [-\frac{\partial f_e(E,\mu,T)}{\partial E}](E-\mu)T_e(E)\,dE\}, \qquad (7b)$$

where $[-\frac{\partial f_e(E,\mu,T)}{\partial E}]$ is a negative derivative of distribution function. From Eq. (7a) it can be seen that $L_0$ is mainly determined by the electronic transmission coefficient $T_e(E)$: the larger $T_e(E)$, the larger $L_0$. On the other hand $L_1$ can be separated into two parts, for positive and negative values of ($E-\mu$), respectively. By doing so we see that $L_1$ is proportional to the difference of $T_e(E)$ in the two parts. In Fig. 3(d), we report the electronic transmission coefficient $T_e$ of *A*-BCNNRs with width $W_A = 5$ at $L_{BN} = 6$ as a function of the electron energy, where the middle line corresponds to the energy position of the chemical potential with respect to the highest figure of merit. It is found that $T_e$ in the region $E<\mu$ decreases abruptly and an energy gap appears with increasing $N$, while $T_e$ at $E>\mu$ varies slowly, leading to $L_0$ decreasing and $L_1$ increasing. Accordingly, the Seebeck coefficient increases with the periodic number.

In Fig. 4, the influence of the length of *h*-BN and graphene on the thermoelectric properties of *A*-BCNNRs is studied. In the left column (a), (c) and (e), we fix the length of *h*-BN $L_{BN}$ and change $L_C$; while in the right column (b), (d) and (f) we fix $L_C$ and change $L_{BN}$. One can see that with the increase of the periodic number $N$, the figure of merit for nanoribbons with different width gradually increases and finally trend to a constant. Compared to the pristine *A*-GNRs, the $ZT$ of metallic nanoribbons is increased about 10~20 times [see the insets of Figs. 4(a) and (b)], while the $ZT$ of semiconducting nanoribbons is enhanced about 1.5~3 times [see the insets of Figs. 4(c-f)]. From these results we can conclude that the thermoelectric properties of $A$-GNRs can be tuned by embedding $h$-BN.

We next investigate the effect of component ratio of $h$-BN on the thermoelectric transport properties in $A$-BCNNRs. Figure 5 shows the figure of merit ratio $ZT/ZT_0$ versus $\gamma$ for nanoribbons with width $W_A = 5$ under different supercell length $L_S$. One can find that $ZT/ZT_0$ at $\gamma = 0$ and 1.0 is small, while in the range of $\gamma$ between 0.2 and 0.8, in particular $\gamma$ near 0.5, $ZT/ZT_0$ attains a relatively large value. The phenomenon can be qualitatively understood from the interface scattering and electronic energy band structure. For the pristine $A$-GNRs, there exist perfect channels of carriers transport. By periodically embedding $h$-BN into the $A$-GNRs, on the one hand, the transport channels are destroyed, resulting in the thermal conductivity and electrical conductance reducing, while on the other hand, a number of minigaps appear in the transmission spectrum. When the value of the component ratio $\gamma$ gets close to 0.5, the width of some minigaps is sufficient to contribute to a high Seebeck coefficient. Moreover, the high Seebeck coefficient and decreased thermal conductivity outweigh the decreased electrical conductance. This implies that generating an effective electronic energy gap is important for improving the thermoelectric performance of materials.

The thermoelectric properties of $Z$-BCNNRs are discussed in Fig. 6 where we show the figure of merit $ZT'$ as a function of the periodic number $N$ for zigzag-edged nanoribbons with width $W_Z = 3$ at a fixed component ratio $\gamma' = 0.5$ ($L'_{BN} = L'_C$), and the inset plot the ratio $ZT'/ZT'_0$, where $ZT'_0$ correspond to the highest figure of merit peak value of $Z$-GNRs. One can find that likewise to the case of $A$-BCNNRs, $ZT'$ gradually increases with the periodic number and ultimately reaches a constant value. The figure of merit $ZT'$ for $Z$-BCNNRs is improved up to 2~3 times compared to that of perfect $Z$-GNRs as shown in the inset. The enhancement of the figure of merit for zigzag-edged nanoribbons is smaller than that of the metallic $A$-GNRs, while is slightly larger than that of the semiconducting $A$-GNRs as is seen by comparing Fig. 6 with Fig. 2. The origin of this behavior can be attributed to the zigzag-edged nanoribbons with higher thermal conductivity than that of the armchair-edged nanoribbons of comparable width. As to the semiconducting $A$-GNRs, because of the energy gap, the thermoelectric figure of merit $ZT_0$ is higher than $ZT_0^{'}$, which outweighs the discrepancy of thermal conductivity, resulting in a $ZT/ZT_0$ falling.

Figure 7 investigates the thermoelectric properties of $Z$-BCNNRs under different component ratio of $h$-BN and graphene. We first fix the length of $h$-BN in the supercell and change the length of graphene [Fig. 7(a)], and then fix the length of graphene and change the length of $h$-BN [Fig. 7(b)]. The insets correspond to the figure of merit ratio $ZT^{'}/ZT_0^{'}$ as a function of periodic number $N$, where $ZT_0^{'}$ is the figure of merit of pristine $Z$-GNRs. One can find that the $ZT^{'}$ for $Z$-BCNNRs increases gradually with increasing the periodic number and finally reaches a constant. The enhanced magnitude, compared to the perfect $Z$-GNRs, is about 2~3 times (see the insets). The enhancement originates from the dramatically decreased thermal conductivity and increased Seebeck coefficient outweighing the decreased electric conductance. In Fig. 7(c) the figure of merit ratio $ZT^{'}/ZT_0^{'}$ versus the component ratio $\gamma'$ for various supercell lengths $L_S^{'}$ is studied. It can be seen that $ZT^{'}/ZT_0^{'}$ in the range of $\gamma'$ from 0.2 to 0.9 exhibits quite larger values and even exceeds 3.0. When $\gamma'$ grows towards 1.0, $ZT^{'}/ZT_0^{'}$ decreases rapidly, while keeping a positive value in the tail because of the embedded $h$-BN.

## IV. CONCLUSIONS

In summary, we have investigated the thermoelectric transport properties of BCNNRs using the NEGF technique. It is found that the thermoelectric figure of merit can be significantly enhanced by periodically embedding $h$-BN into GNRs. Compared to the pristine GNRs, the $ZT$ of $A$-BCNNRs with the width index of $3p+2$ is enhanced up to 10~20 times, while $ZT$ of nanoribbons with other widths is enhanced about 1.5~3 times; as to zigzag-edge, the figure of merit for BCNNRs is improved 2~3 times. The strong enhancement can be justified by the combined increase in the Seebeck coefficient and decrease in the thermal conductivity outweighing the decrease in the electrical conductance. In addition, the effect of component ratio of *h*-BN on the thermoelectric properties is discussed. It is shown that the figure of merit in the case of $\gamma$ ($\gamma'$) nearing 0.5 exhibits a larger value. Our investigations demonstrate that the hybrid nanostructure composed of alternating *h*-BN and graphene is a very promising material for building highly efficient thermoelectric devices.

## ACKNOWLEDGEMENTS

We acknowledge financial support from Spanish MICINN (*FIS2011-65702-C02-01* and *PIB2010US-00652*), ACI-Promociona (*ACI2009-1036*), Grupos Consolidados UPV/EHU del Gobierno Vasco (*IT-319-07*), Consolider nanoTHERM (*Grant No. CSD2010-00044*), the European Research Council Advanced Grant DYNamo (*ERC-2010-AdG -Proposal No. 267374*), and also the National Natural Science Foundation of China (*Grant Nos. 51176161, 51006086, 11074213 and 10774127*).

## Figure captions

**Fig. 1.** (Color online) Schematics of (a) and (b) $A$-BCNNRs, and (c) and (d) $Z$-BCNNRs. The left and right leads made of GNRs have perfect periodicity along the ribbon axis. The central scattering region is formed by periodic alternating $h$-BN and graphene. The dotted pane denotes a supercell in which $L_{BN}$ ($L'_{BN}$) and $L_C$ ($L'_C$) separately represent the lengths of $h$-BN and graphene for armchair- and zigzag-edged nanoribbons in the longitudinal direction, while $W_A$ and $W_Z$ are used to denote the width of the nanoribbons, respectively. Note that the length of $h$-BN in (a) and (c) equals to the length of graphene, while in (b) and (d) their lengths are different.

**Fig. 2.** (Color online) Thermoelectric figure of merit $ZT$ for $A$-BCNNRs with different widths: (a) $W_A = 5$, (b) $W_A = 6$ and (c) $W_A = 7$, while the insets plot the ratio $ZT/ZT_0$ as a function of periodic number $N$ for the nanoribbons, where $ZT_0$ corresponds to the result of perfect $A$-GNRs. The component ratio $\gamma = 0.5$ (i.e., $L_{BN} = L_C$), and the temperature $T = 300K$.

**Fig. 3.** (Color online) (a) Electrical conductance ratio $\sigma/\sigma_0$, (b) thermal conductivity ratio $\kappa/\kappa_0$ and (c) Seebeck coefficient ratio $S/S_0$ as a function of periodic number $N$ for $A$-BCNNRs with width $W_A = 5$, where $\sigma_0$, $\kappa_0$ and $S_0$ correspond to pristine $A$-GNRs; the insets show the results of nanoribbons

with width $W_A = 6$. (d) Transmission coefficient $T_e$ versus electronic energy $E$ for A-BCNNRs with $W_A = 5$ at $L_{BN} = 6$, where $t$ is the hopping energy between two nearest-neighbor atoms in graphene. The other parameters are the same as in Fig. 2.

**Fig. 4.** (Color online) Thermoelectric figure of merit $ZT$ for A-BCNNRs as a function of periodic number $N$. In the left column (a), (c) and (e) $L_{BN} = 6$; while in the right column (b), (d) and (f) $L_C = 6$. The insets plot the ratio $ZT/ZT_0$ of the nanoribbons. The other parameters are set the same as in Fig. 2.

**Fig. 5.** (Color online) Thermoelectric figure of merit ratio $ZT/ZT_0$ of A-BCNNRs with width $W_A = 5$ versus component ratio $\gamma$, of which the periodic number $N = 10$ and temperature $T = 300K$.

**Fig. 6.** (Color online) Thermoelectric figure of merit $ZT'$ for Z-BCNNRs at room temperature, and the inset plot the ratio $ZT'/ZT_0'$ versus periodic number $N$, where $ZT_0'$ is the result of pristine Z-GNRs. Here we take $\gamma' = 0.5$ ($L_{BN}' = L_C'$) and $W_Z = 3$.

**Fig. 7.** (Color online) Thermoelectric figure of merit for Z-BCNNRs versus periodic number $N$ under various component ratio. In (a) $L_{BN}' = 6$ and (b) $L_C' = 6$. The insets represent the ratio $ZT'/ZT_0'$ of the nanoribbons. (c) Figure of merit ratio versus component ratio $\gamma'$ for different supercell length $L_S'$. The other parameters are the same as in Fig. 6.

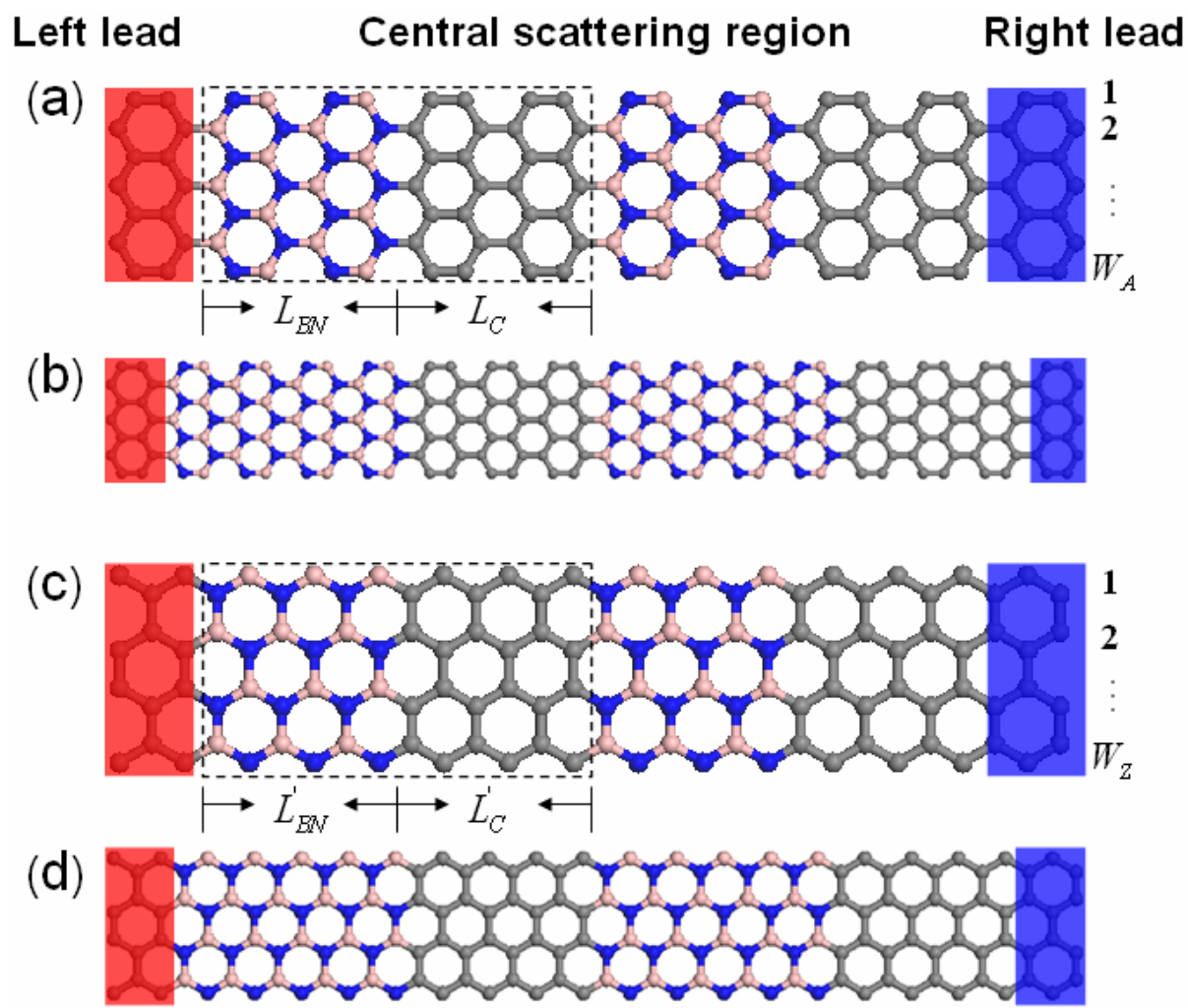

Fig. 1

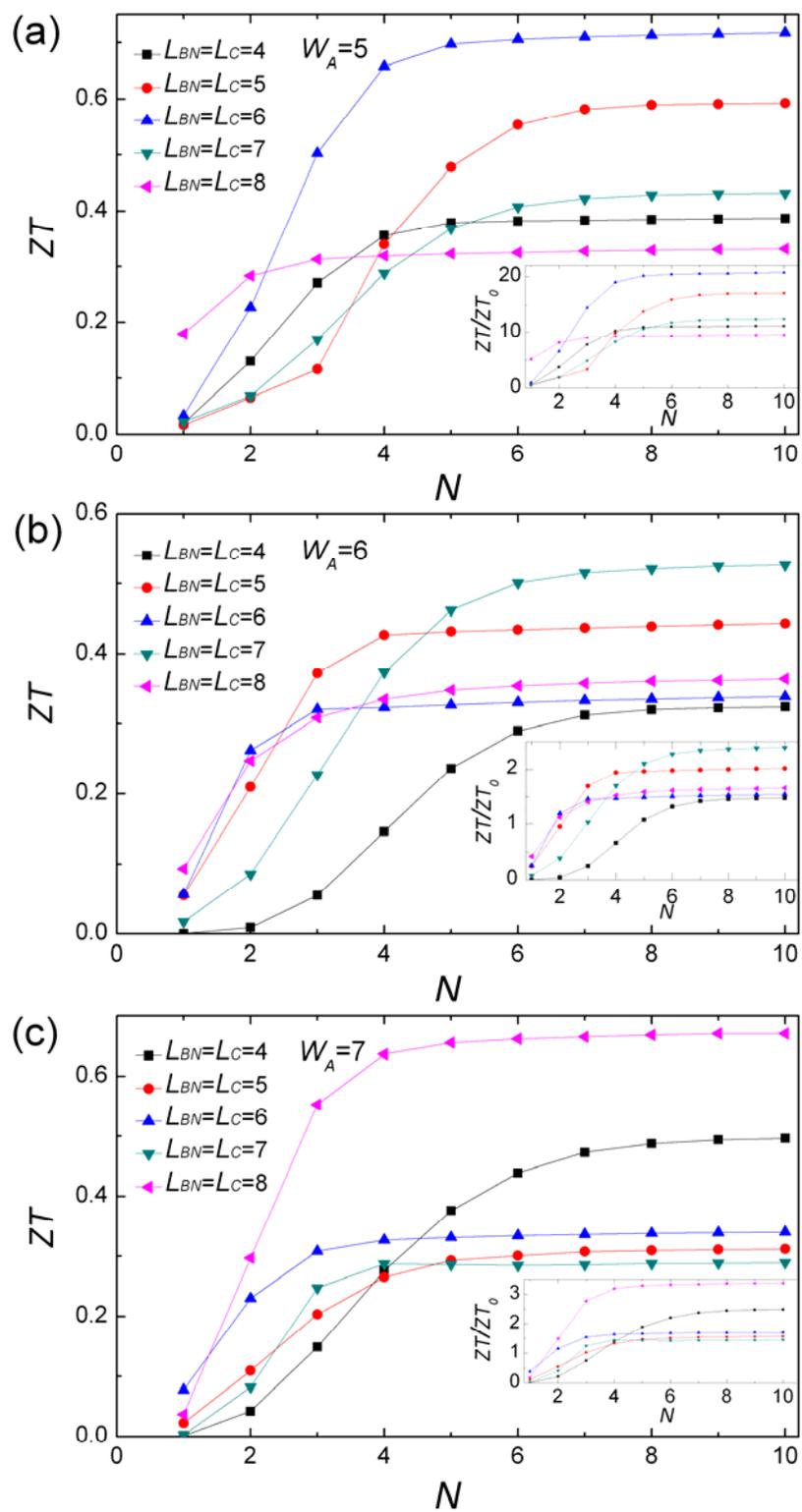

Fig. 2

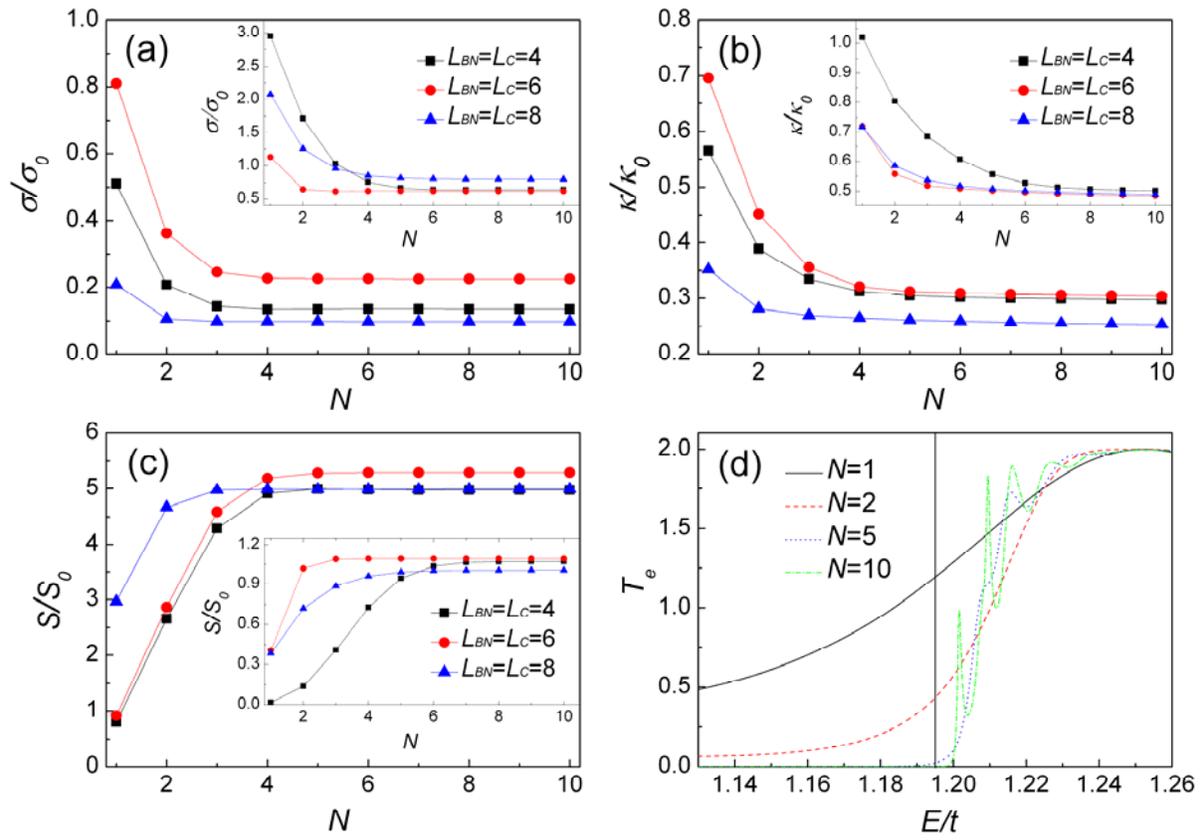

Fig. 3

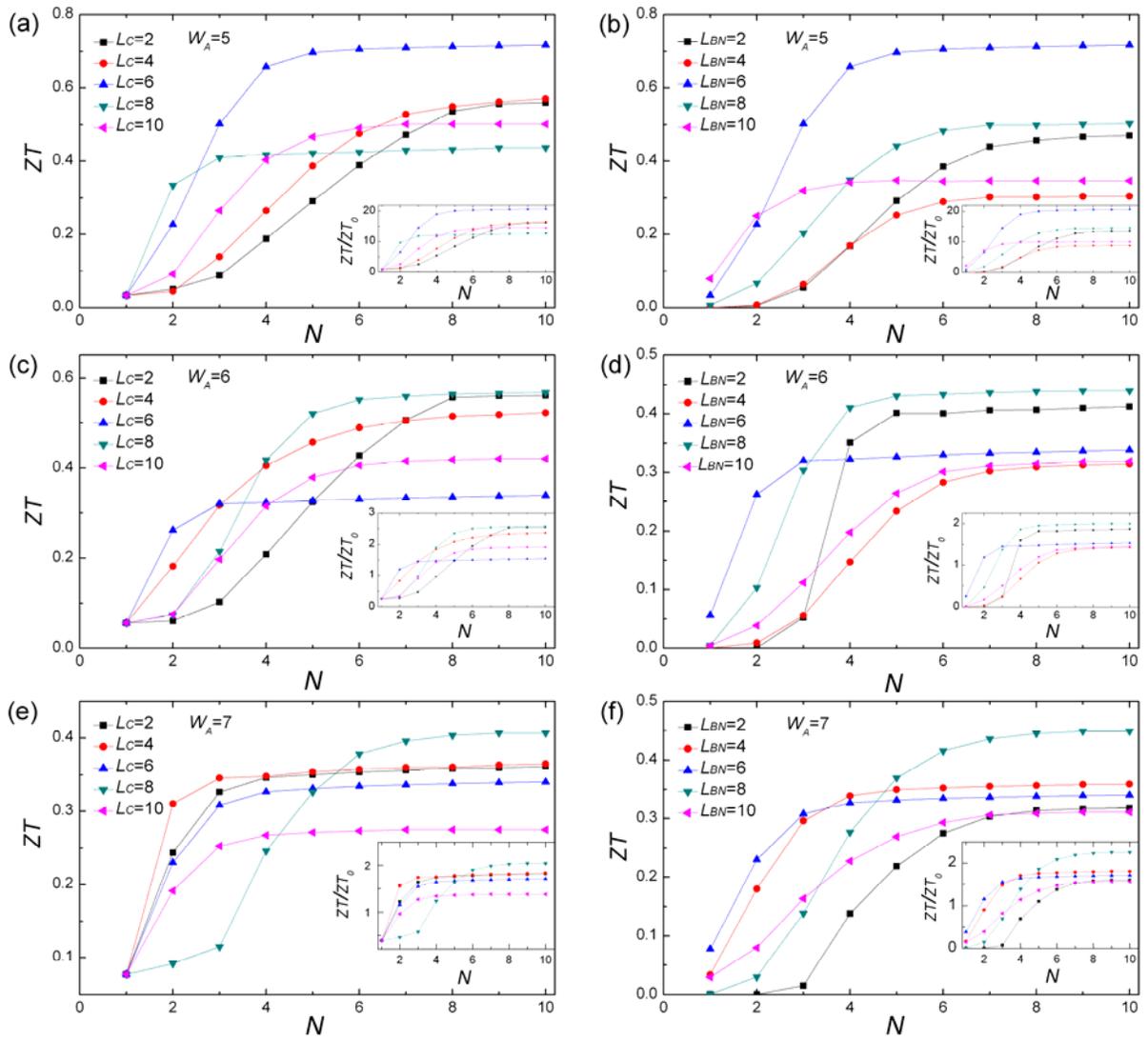

Fig. 4

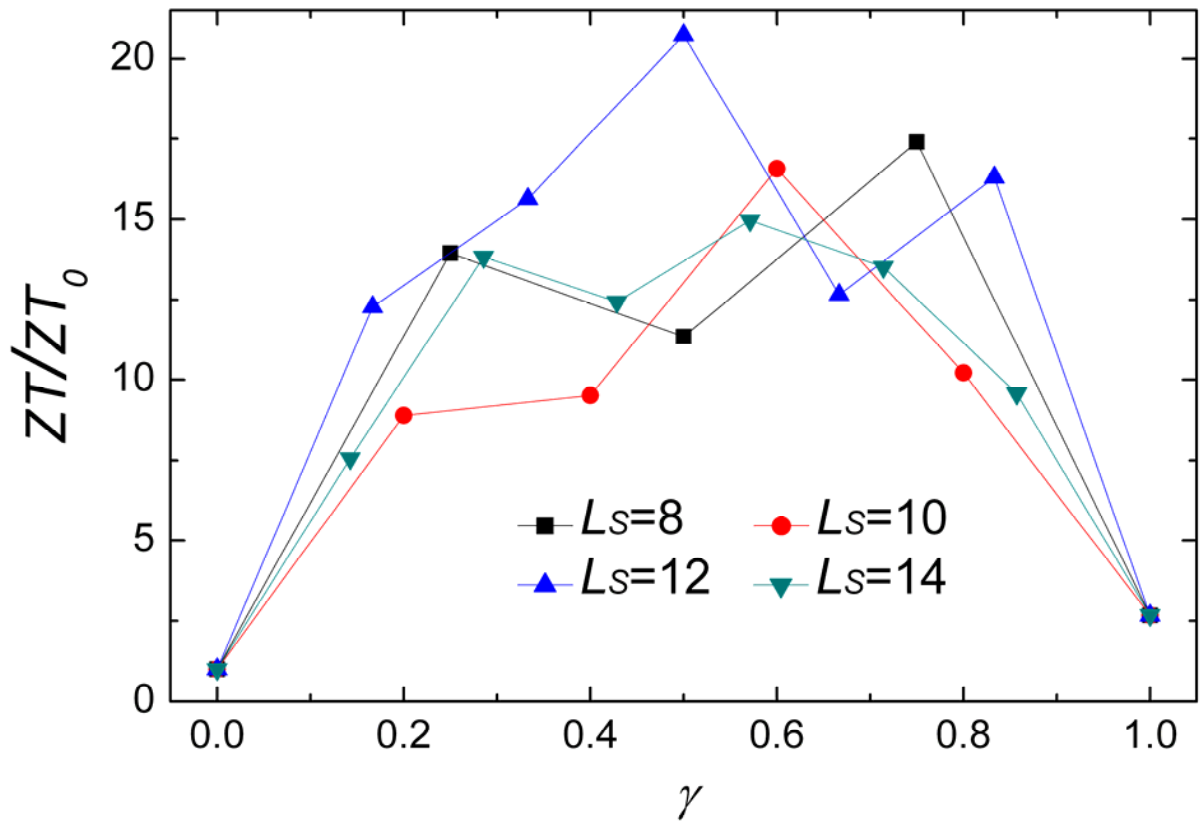

Fig. 5

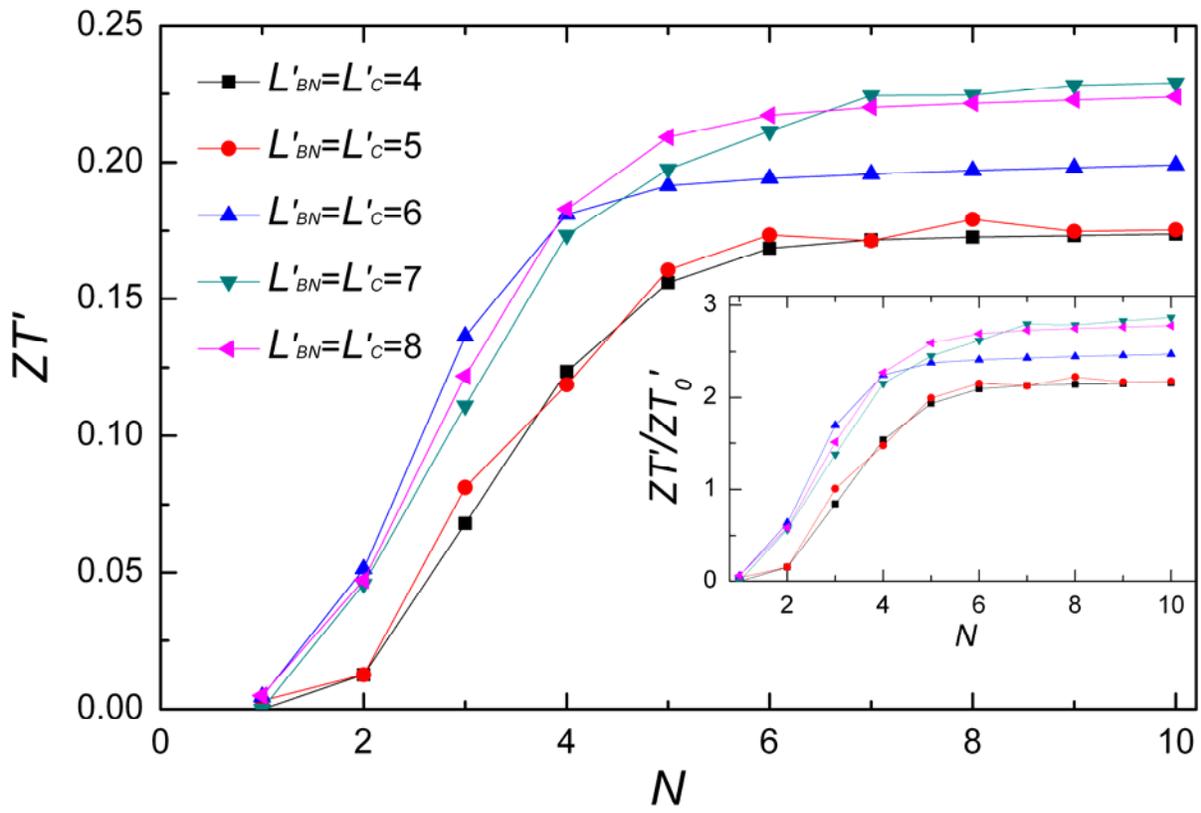

Fig. 6

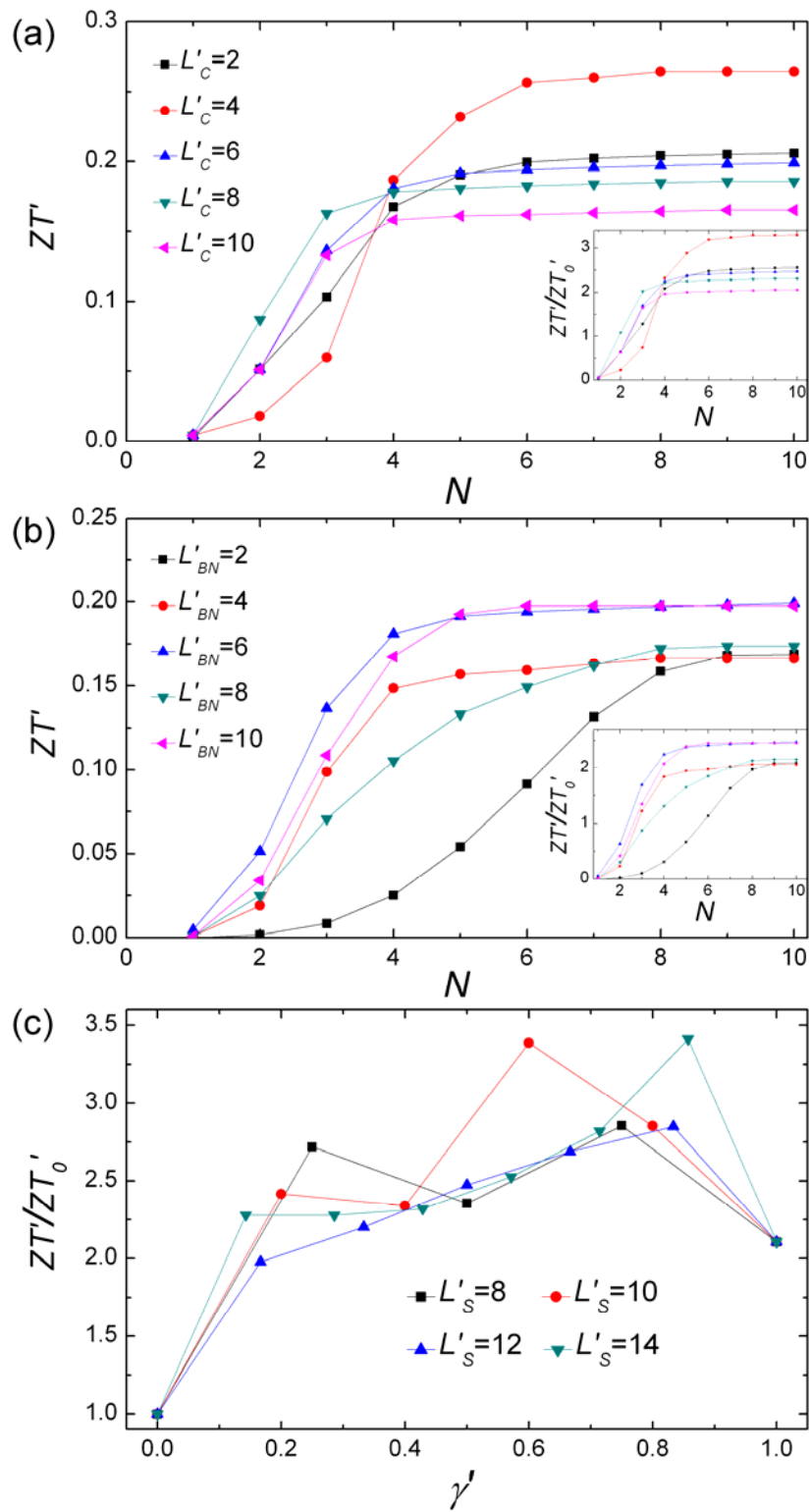

Fig. 7